\documentclass[superscriptaddress, reprint, prl, showpacs]{revtex4-1}

\usepackage[english]{babel}

\usepackage{amsmath}
\usepackage{amsthm}
\usepackage{amssymb}

\usepackage{hyperref}
\usepackage{graphicx}

\begin{document}


\title{Electric-field-induced pyroelectric order and localization of the confined electrons in LaAlO$_3$/SrTiO$_3$ heterostructures}

\author{M. R\"ossle}
\affiliation{University of Fribourg, Department of Physics and Fribourg Center for Nanomaterials, Chemin du Mus\'ee 3, CH-1700 Fribourg, Switzerland} 
\author{K.W. Kim}
\affiliation{University of Fribourg, Department of Physics and Fribourg Center for Nanomaterials, Chemin du Mus\'ee 3, CH-1700 Fribourg, Switzerland}
\affiliation{Department of Physics, Chungbuk National University, Cheongju 361-763, Korea} 
\author{A. Dubroka} 
\affiliation{University of Fribourg, Department of Physics and Fribourg Center for Nanomaterials, Chemin du Mus\'ee 3, CH-1700 Fribourg, Switzerland} 
\affiliation{Department of Condensed Matter Physics, Faculty of Science, Masaryk University and Central European Institute of Technology, Kotl\'a\u rsk\'a 2, CZ-61137 Brno, Czech Republic}
\author{P. Marsik}
\affiliation{University of Fribourg, Department of Physics and Fribourg Center for Nanomaterials, Chemin du Mus\'ee 3, CH-1700 Fribourg, Switzerland} 
\author{C.N. Wang} 
\affiliation{University of Fribourg, Department of Physics and Fribourg Center for Nanomaterials, Chemin du Mus\'ee 3, CH-1700 Fribourg, Switzerland} 
\author{R. Jany}
\affiliation{Experimental Physics VI, Center for Electronic Correlations and Magnetism, Institute of Physics, University of Augsburg, D-86135 Augsburg, Germany}
\author{C. Richter}
\affiliation{Experimental Physics VI, Center for Electronic Correlations and Magnetism, Institute of Physics, University of Augsburg, D-86135 Augsburg, Germany}
\affiliation{Max-Planck-Institut f\"ur Festk\"orperforschung, Heisenbergstrasse 1, D-70569 Stuttgart, Germany}
\author{J. Mannhart}
\affiliation{Max-Planck-Institut f\"ur Festk\"orperforschung, Heisenbergstrasse 1, D-70569 Stuttgart, Germany}
\author{C.W. Schneider} 
\affiliation{Paul Scherrer Institut, CH-5232 Villigen, Switzerland}
\author{A. Frano} 
\affiliation{Max-Planck-Institut f\"ur Festk\"orperforschung, Heisenbergstrasse 1, D-70569 Stuttgart, Germany}
\affiliation{Helmholtz-Zentrum Berlin f\"ur Materialien und Energie, Albert-Einstein-Strasse 15, D-12489 Berlin, Germany}
\author{P. Wochner} 
\affiliation{Max-Planck-Institut f\"ur Intelligente Systeme, Heisenbergstrasse 3, D-70569 Stuttgart, Germany}
\author{Y. Lu}
\affiliation{Max-Planck-Institut f\"ur Festk\"orperforschung, Heisenbergstrasse 1, D-70569 Stuttgart, Germany}
\author{B. Keimer} 
\affiliation{Max-Planck-Institut f\"ur Festk\"orperforschung, Heisenbergstrasse 1, D-70569 Stuttgart, Germany}
\author{D. K. Shukla}
\affiliation{Deutsches Elektronen-Synchrotron DESY, Notkestrasse 85, D-22603 Hamburg, Germany} 
\author{J. Strempfer} 
\affiliation{Deutsches Elektronen-Synchrotron DESY, Notkestrasse 85, D-22603 Hamburg, Germany} 
\author{C. Bernhard}
\email{christian.bernhard@unifr.ch}
\affiliation{University of Fribourg, Department of Physics and Fribourg Center for Nanomaterials, Chemin du Mus\'ee 3, CH-1700 Fribourg, Switzerland} 
 
\begin{abstract}
With infrared ellipsometry, x-ray diffraction, and electric transport measurements we investigated the electric-field-effect on the confined electrons at the LaAlO$_3$/SrTiO$_3$ interface. We obtained evidence that the localization of the electrons at low temperature and negative gate voltage is induced, or at least strongly enhanced, by a pyroelectric phase transition in SrTiO$_3$ which strongly reduces the lattice polarizability and the subsequent Coulomb screening. In particular, we show that the charge localisation and the polar order of SrTiO$_3$ both develop below $\sim50$~K and exhibit similar, unipolar hysteresis loops as a function of the gate voltage. Our findings suggest that the pyroelectric order also plays an important role in the quantum phase transition at very low temperatures where superconductivity is suppressed by an electric field. 
\end{abstract}


\maketitle

\begin{figure}[!ht]
  \includegraphics[width = \columnwidth]{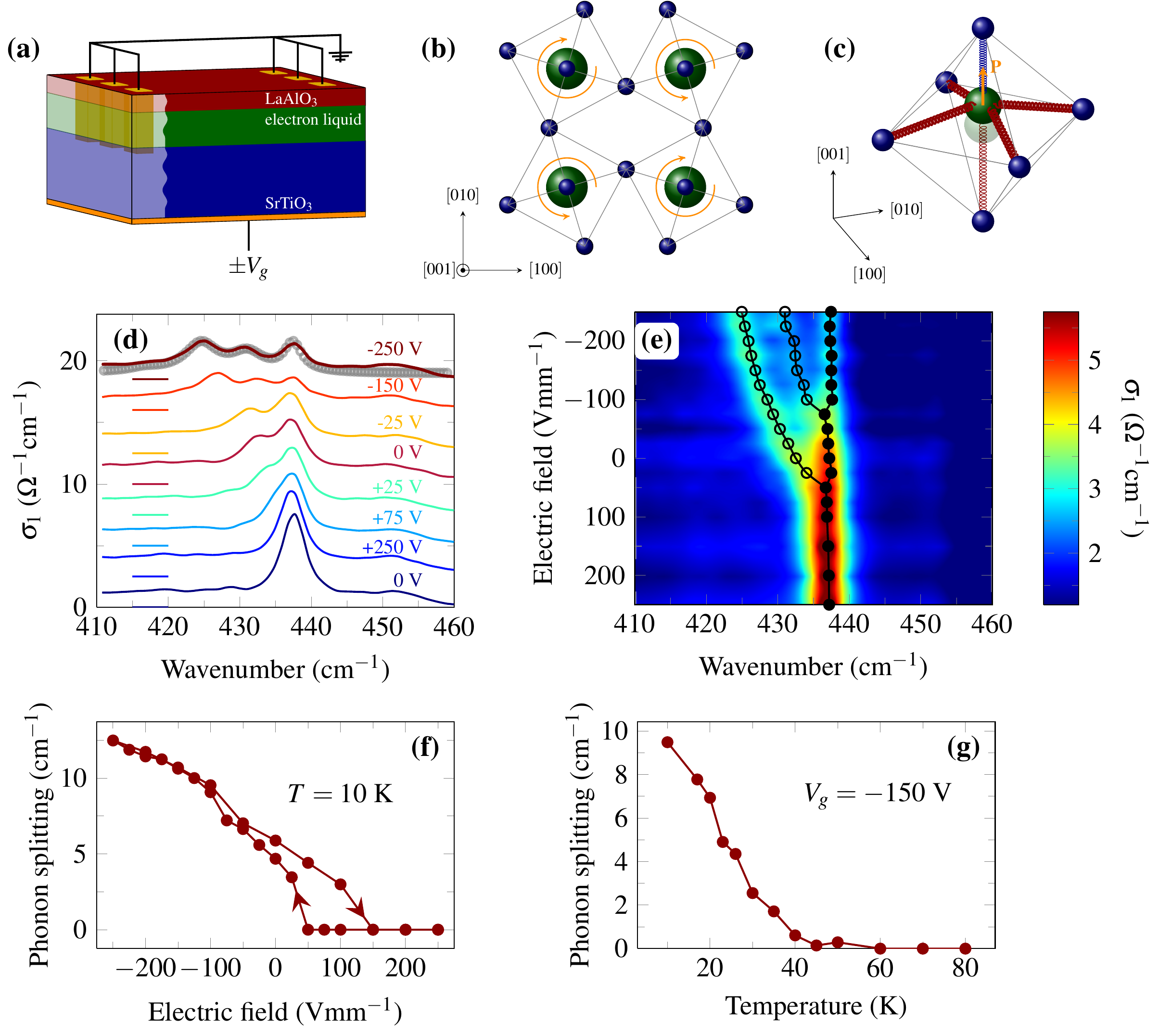}
  \caption{\label{fig1} (a) Schematics of the LAO/STO heterostructure and the electrodes used to apply back-gate voltages up to $\pm250$~V. (b) Anti-phase rotation of the TiO$_6$ octahedra in tetragonal SrTiO$_3$ that gives rise to the infrared-active $R$-mode at 438~cm$^{-1}$. (c) Sketch of the dipolar displacement of the Ti-ions along the $[001]_c$-axis showing the effect on the length and the eigenfrequency of the TiO bonds as discussed in the text. (d) and (e) Voltage-dependence of the infrared spectra at 10~K showing the anomalous splitting and softening of the $R$-mode. Solid lines in (d) quantify the $y$-shifts added for clarity. Solid circles in (e) mark the peak at 438~cm$^{-1}$ due to paraelectric STO, open circles the two new peaks from the pyroelectric STO layer. (f) Electric field dependence of the maximal splitting of the $R$-mode. (g) Temperature dependence of the $R$-mode splitting.}
\end{figure}

The highly mobile electrons at the interface between TiO$_2$ terminated SrTiO$_3$ (STO) substrates and thin LaAlO$_3$ (LAO) layers are the subject of intense research efforts \cite{Ohtomo2004}. It is meanwhile well established that the confined electrons at the LAO/STO interface develop rather suddenly as the LAO layer reaches a minimal thickness of $n = 4$ unit cells \cite{Thiel2006}. While the mixing of cations \cite{Willmott2007} or oxygen vacancies \cite{Kalabukhov2007} at the interface may also contribute, it is widely believed that the confined electrons originate from a charge transfer from the top surface of the LAO layer to the LAO/STO interface \cite{Nakagawa2006, Thiel2006}. The latter is a consequence of a so-called polar catastrophe that arises because the polar layer stacking yields an electrostatic potential that increases with the LAO layer thickness \cite{Thiel2006}. It has been shown that the conductivity of the confined electrons can be readily modified with a gate voltage which enables the design of efficient field-effect devices \cite{Thiel2006}. For LAO/STO heterostructures with $n = 3$ that are still insulating in their initial state, it has been demonstrated that a metallic state can be induced with moderate electric fields \cite{Thiel2006} that are readily produced with gate electrodes as shown in Figure~\ref{fig1}(a). Even the biased tip of an AFM has been used to write and erase conducting pathways \cite{Cen2009}. For samples with $n \geq 4$, the conductivity of the confined electrons can also be modified with a gate voltage, albeit a field-induced metal-to-insulator transition occurs only at low temperature \cite{Liao2011}. The confined electrons even become superconducting below a transition temperature of $T_c \sim 0.2 - 0.4$~K \cite{Reyren2007}. Notably, this two-dimensional superconducting state can coexist with a ferromagnetic order that develops at much higher temperature \cite{Bert2011, Dikin2011}. The superconductivity can be entirely suppressed with an electric field which induces a quantum phase transition toward a localized state \cite{Caviglia2008}. The mechanism by which the external electric field induces this localization, e.g. whether it changes the concentration or the interaction strength and mobility of the confined electrons, is still under discussion \cite{Liao2011, Bell2009, Caviglia2010, Biscaras2012}.

In the following we present combined infrared ellipsometry, x-ray diffraction and electric transport measurements which reveal that a pyroelectric phase transition in SrTiO$_3$ plays a dominant role in the electric-field-induced localization of the confined charge carriers. 

The LAO/STO heterostructures have been prepared similarly as described in Ref.~\cite{Thiel2006}. A sketch of the contact arrangement and the polarity of the applied gate voltage is displayed in Fig.~\ref{fig1}(a). A SrTi$^{18}$O$_3$ single crystal which serves as a ferroelectric reference sample has been obtained by placing a SrTiO$_3$ substrate (purchased from Crystec) for one week at 1100$^{\circ}$~C in an oxygen atmosphere with an $^{18}$O isotope content of 92\%. This resulted in the exchange of $^{16}$O with $^{18}$O into a depth of approximately 100~$\mu$m. The isotope content has been verified by Raman spectroscopy to be close to 90\% based on the redshift of the phonon mode at 683~cm$^{-1}$ to 650~cm$^{-1}$.
 
The infrared ellipsometry experiments have been performed with a home-built setup that is attached to a commercial fast Fourier-Transform interferometer (Bruker 113v) as described in Ref. \cite{Bernhard2004}. For the modeling of the ellipsometry data we used the Woollam VASE software \cite{Woollam}. The x-ray diffraction experiments have been conducted at beamline P09/PETRA III at DESY, Hamburg, Germany, at a photon energy of 9.8~keV. The sample was cooled to 10~K using a closed cycle cryostat mounted on a 6-circle diffractometer. The reciprocal space was mapped using an APD detector with a high dynamic range. For the two-point resistivity measurements we used a Quantum Design PPMS system in combination with an external gate voltage source (Tennelec TC952) and a Keithley 2602A source meter.

First we present the infrared ellipsometry study of the electric-field-effect on the dielectric response of a LAO/STO heterostructure with an average LAO layer thickness of about 5.5 unit cells. Initially, we wanted to search for changes of the mobility and the concentration depth profile of the confined electrons following the approach of Ref.~\cite{Dubroka2010}. To our surprise, we observed some additional, pronounced features which arise from an anomaly of an infrared-active phonon mode that are characteristic of an electric-field-induced, unipolar electric (or pyroelectric) order in the STO layer adjacent to the LAO/STO interface. The infrared spectra at 10~K showing the electric-field-effect of the relevant STO phonon mode are displayed in Figs.~\ref{fig1}(d) and \ref{fig1}(e). This so-called $R$-mode at 438~cm$^{-1}$ develops below the antiferro-distortive phase transition at 105~K at which the crystal symmetry changes from cubic to tetragonal \cite{Lytle1964, Loetzsch2010}. As indicated in Fig.~\ref{fig1}(b), the $R$-mode involves an antiphase rotation of the neighbouring TiO$_6$ octahedra. The mode is located at the $R$-point of the cubic Brillouin-zone but becomes weakly infrared-active in the tetragonal state due to the back-folding of the Brillouin-zone \cite{Fleury1968a, Petzelt2001}. The sketch in Fig.~\ref{fig1}(c) shows how the $R$-mode of STO is affected by the formation of a dipolar, electric order at which the Ti ions are displaced from the centers of the TiO$_6$ octahedra. Shown is a (exaggerated) displacement along the $[001]_c$ direction (in cubic notation), the alternative case of a displacement along the $[110]_c$ or $[111]_c$ direction is discussed in Ref. \cite{SOM}. The colored springs indicate the red- and blue-shift of the eigenfrequency of the Ti-O bonds that are caused by the respective increase or decrease of the bond lengths. This sketch illustrates that the polar displacement yields an anisotropic softening and thus a splitting of the $R$-mode. The larger softening occurs for the component with the light polarization perpendicular to the polar displacement. This kind of softening and splitting of the $R$-mode is indeed observed in Figs.~\ref{fig1}(d) and \ref{fig1}(e). The $R$-mode is hardly affected as $V_g$ is initially increased from 0 to $+250$~V whereas it develops two side peaks at lower frequency as $V_g$ is subsequently decreased to $-250$~V. The two softer modes can be assigned to a pyroelectric STO layer in the vicinity of the LAO/STO interface and the original peak at 438~cm$^{-1}$ to the paraelectric bulk of the STO substrate. The continuous spectral weight transfer between the original and the two new $R$-modes suggests that the thickness of the pyroelectric layer increases continuously and reaches a value of $d_{\mathrm{polar}}= 1~\mu$m at $V_g = -250$~V. The thick grey line of circles in Fig.~\ref{fig1}(d) shows the calculation for $d_{\mathrm{polar}} = 1~\mu$m which describes best the spectrum at $V_g = -250$~V. All parameters, except for the position, the width and the relative strength of the softened $R$-modes were obtained here from the spectrum at $+250$~V as outlined in Ref. \cite{SOM}. Figure~\ref{fig1}(f) shows how the splitting of the phonon mode evolves with $V_g$. It highlights the unipolar nature of this phenomenon and reveals its hysteretic behavior. Figure~\ref{fig1}(g) displays the $T$-dependence of the mode splitting at $V_g = -150$~V and shows that the pyroelectric order develops below about 50~K, i.e. near the avoided phase transition of the incipient ferroelectric SrTiO$_3$ \cite{Tosatti1994}.

\begin{figure}[!ht]
  \includegraphics[width = \columnwidth]{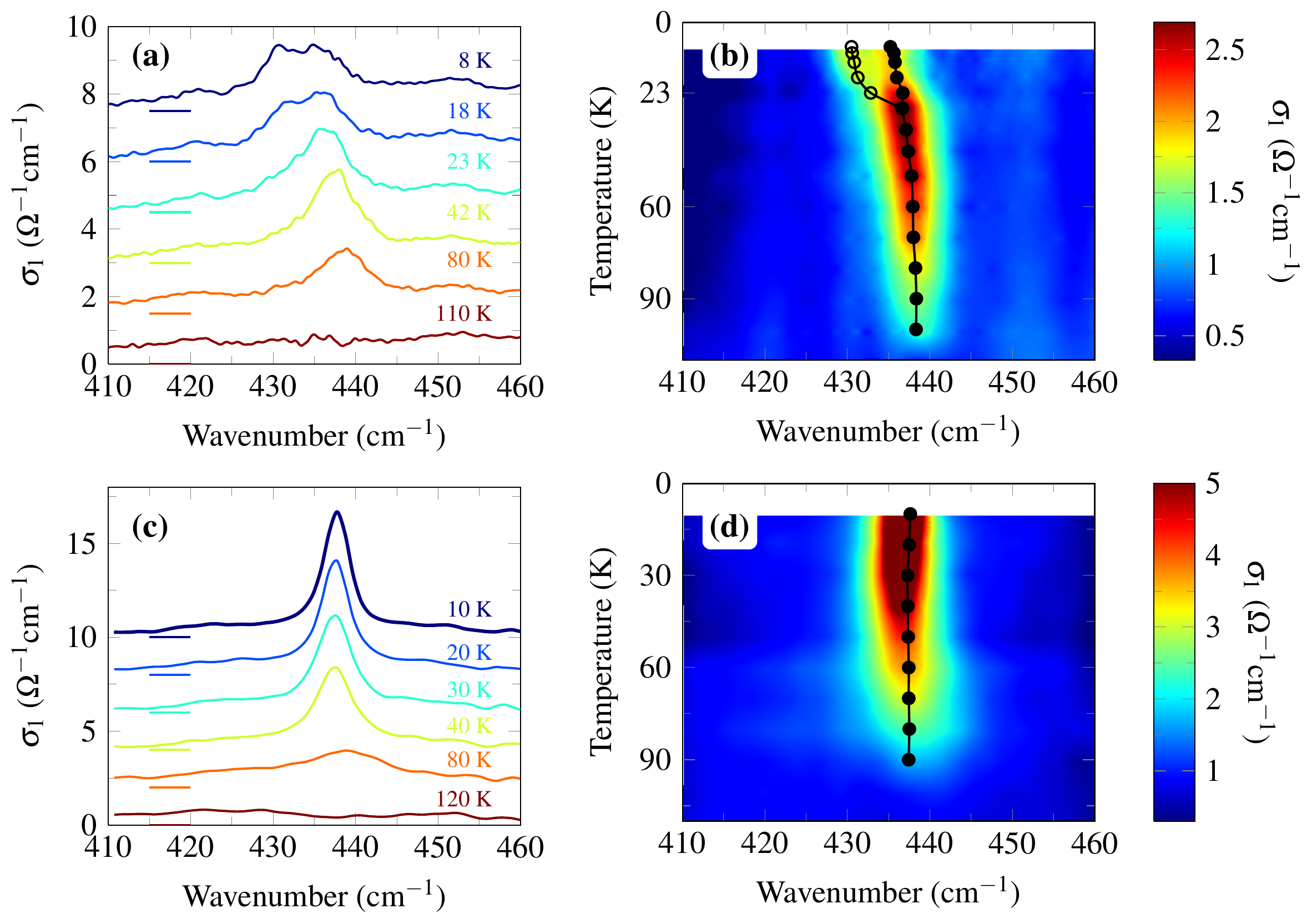}
  \caption{\label{fig2} (a) and (b) Infrared spectra showing the temperature dependence of the $R$-mode in a SrTi$^{18}$O$_3$ single crystal with a ferroelectric transition at $T^{\textrm{Curie}} = 23$~K. (c) and (d) Corresponding spectra for a SrTi$^{16}$O$_3$ crystal which remains in the quantum paraelectric state.}
\end{figure}

Figures~\ref{fig2}(a) and \ref{fig2}(b) show that a corresponding, characteristic splitting and softening of the $R$-mode occurs for a SrTi$^{18}$O$_3$ single crystal which exhibits a well-known ferroelectric transition at $T_{\mathrm{Curie}} \approx 23$~K \cite{Itoh1999, Yamaguchi2001}. As compared to SrTi$^{16}$O$_3$, which remains in a quantum paraelectric state, the ferroelectric phase is stabilized by the isotope-mass-induced reduction of the quantum lattice fluctuations \cite{Muller1979}. Figures \ref{fig2}(c) and \ref{fig2}(d) confirm that the $R$-mode anomaly is entirely absent for SrTi$^{16}$O$_3$ for which an electric field in excess 400~V/mm would be required to induce such a ferroelectric polarization \cite{Hemberger1995}. 

\begin{figure}[!ht]
  \includegraphics[width = \columnwidth]{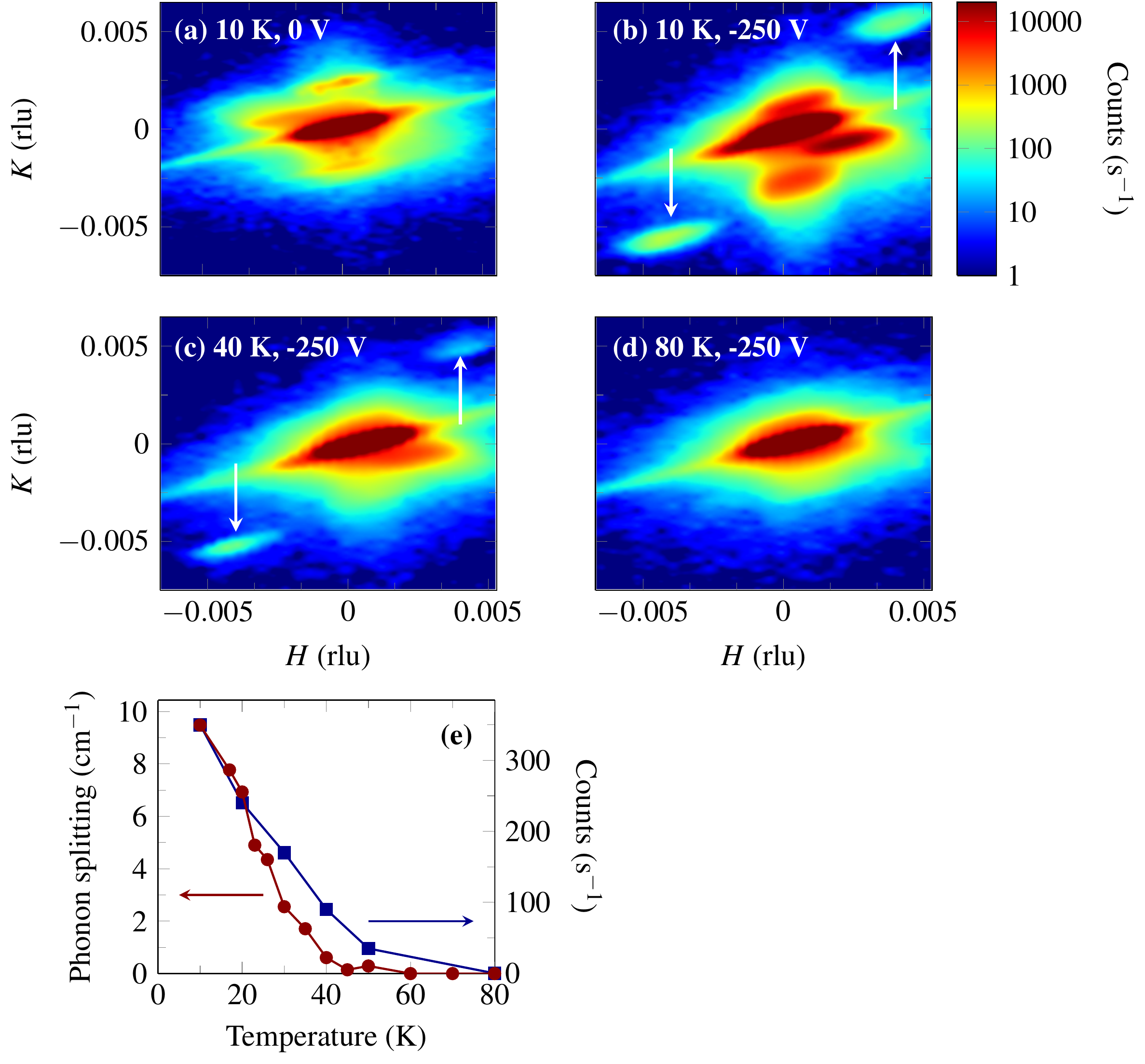}
  \caption{\label{fig3}Field-induced satellite Bragg peaks in the pyroelectric domain state of LAO/STO. (a) and (b) Electric field dependence of the $k$-space map around the (002) Bragg peak of SrTiO$_3$ at $T = 10$~K. White arrows mark two new satellite peaks that appear at $V_g = -250$~V at the (-0.004 -0.005 2) and (0.004 0.005 2) positions. (c) to (e) Temperature dependence of the satellite peaks in comparison with the splitting of the $R$-mode.}
\end{figure}

The signatures of a domain state due to the polar electric order in LAO/STO have also been observed with high resolution x-ray diffraction measurements \cite{Strempfer2012}. Figures~\ref{fig3}(a)-(d) display $k$-space maps in the vicinity of the $(002)$ Bragg peak of SrTiO$_3$ at different gate voltages and temperatures. It reveals that two new satellite peaks appear at $T = 10$~K and $V_g = -250$~V. Their spacing of $\Delta Q = 0.0063$~reciprocal lattice units (rlu) with respect to the STO $(002)$ Bragg peak signifies a modulation of the structural parameters of the interfacial STO layer in the direction parallel to the LAO/STO interface with a period of $\sim 60$~nm. The circumstance that only two satellites are observed suggests that this modulation has a preferred in-plane direction along $[110]_c$. It was previously shown \cite{Streiffer2002, Fong2004} that such superlattice peaks typically occur in ferroelectric thin films where they arise from anti-phase stripe domains with an alternating polarization direction. In our LAO/STO heterostructures the normal component of the unipolar electric order likely remains parallel to the applied electric field. The satellite peaks still may be due to the alternation of the direction of an in-plane component that arises from a tilting of the electric polarization or else by an amplitude modulation of the normal component. Irrespective of the unknown details of the domain structure, the mere observation of these satellite peaks confirms that an electric displacement with a stripe domain pattern is induced at $V_g = -250$~V. The observed domain size of $\sim 60$~nm at a layer thickness of $1~\mu$m is roughly consistent with the trend reported in Ref. \cite{Fong2004}. Figures~\ref{fig3}(c) to \ref{fig3}(e) show the temperature dependence of the satellite peaks which disappear around 50~K, similar to the splitting of the $R$-mode in the infrared spectra.

Next we discuss the origin of the unipolar electric order of the interfacial SrTiO$_3$ layer and its relationship to the metal-to-insulator transition of the confined electrons. SrTiO$_3$ is an incipient ferroelectric material for which the correlation length starts to diverge below $\sim 50$~K, but for which long range order is inhibited by the quantum lattice fluctuations \cite{Tosatti1994, Muller1979}. Despite the large thickness of the pyroelectric layer of $\sim 1~\mu$m at $-250$~V, the polar order thus may be governed by the properties of the LAO/STO interface. A built-in electric field that is pointing toward the STO substrate likely plays an important role. It arises from the interfacial electronic and structural reconstruction and yields the band-bending and the confining potential of the interfacial electrons. This local field can explain the unipolar nature of the electric order and also that its onset occurs already at a positive gate voltage (see Fig. 1(f)). A corresponding lattice polarization in the first few STO monolayers has been predicted \cite{Pentcheva2009, Schwingenschloegl2009} and observed in a recent surface x-ray diffraction (SXRD) study \cite{Pauli2011}. The lattice distortion and the underlying local electric field increase as the LAO thickness grows from $n = 2$ to 5. We found indeed that the unipolar field effect on the $R$-mode is significantly smaller in a LAO/STO heterostructure with $n = 3$ and is entirely absent in a control device with a thin Ti electrode that was directly deposited on a SrTiO$_3$ substrate. Our observations thus argue for an important role of the electronic and structural reconstruction at the LAO/STO interface.

It seems unlikely that the confined electrons are playing an active role in the pyroelectric phase transition. The polarity of the electric order yields a negative ionic charge density at the LAO/STO interface for which the interfacial electrons cannot stabilize the order by screening the external electric stray field. On the other hand, the field-induced metal-to-insulator transition of the confined electrons may well be triggered or even driven by the development of the pyroelectric order. The latter reduces the polarizability of the lattice and thus its ability to screen defects that scatter and eventually localize the confined electrons. The polarization also increases the confining potential and thus may enhance the localization effects underlying the metal-to-insulator transition.

\begin{figure}[!ht]
  \includegraphics[width = \columnwidth]{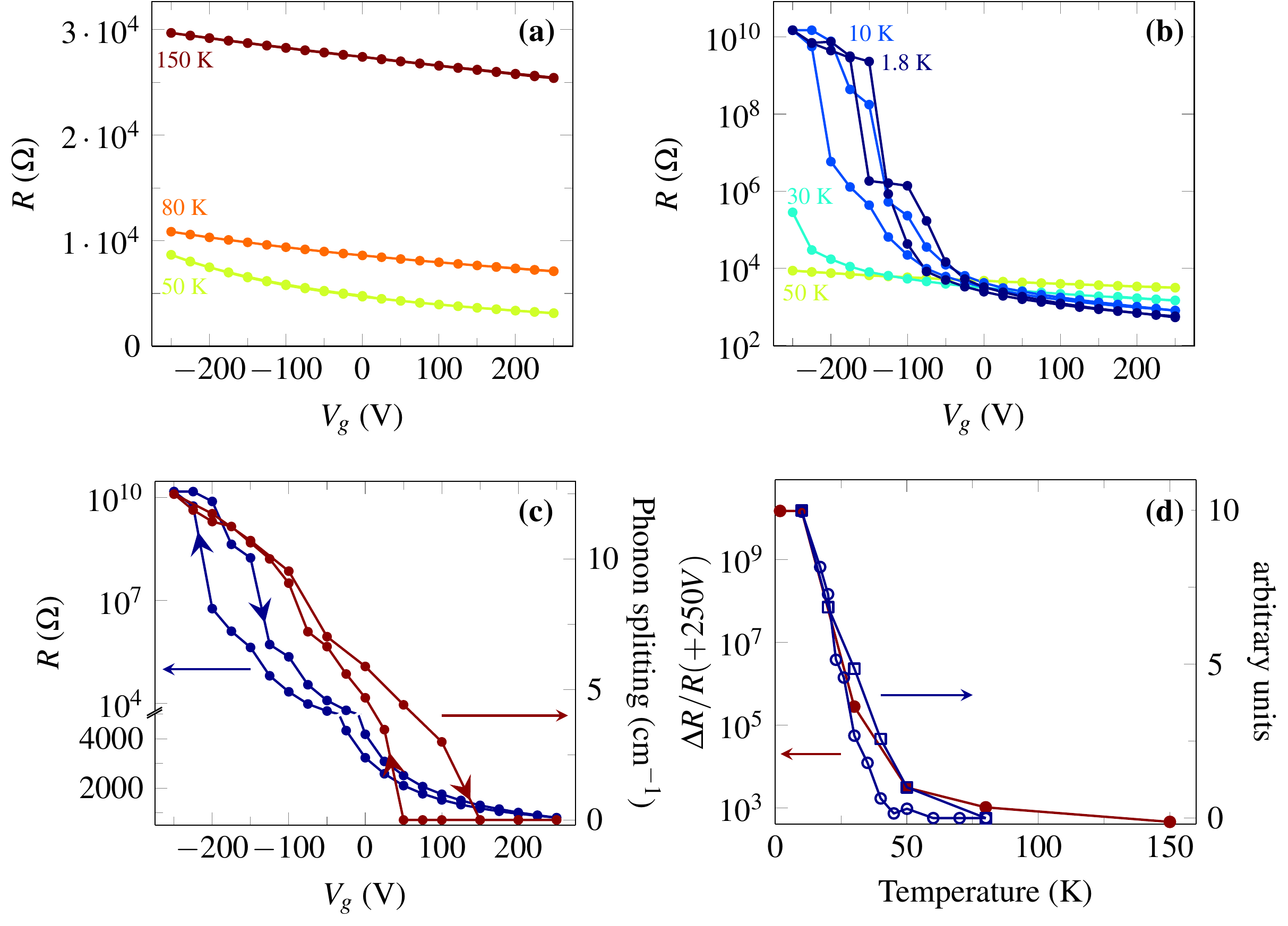}
  \caption{\label{fig4} Gate voltage dependence of the resistance as determined from a two point measurement. (a) and (b) $R-V$ curves showing the change from a metallic regime with a weak, linear dependence at 80 and 150~K to the localisation regime at low temperature where $R$ increases steeply towards negative $V_g$ and exhibits a hysteretic behaviour. (c) Comparison of the voltage loops of the resistivity at 10~K and the splitting of the $R$-mode. (d) Temperature dependence of the field-induced increase in resistance, $R(-250~\mathrm{V})-R(+250~\mathrm{V})/R(+250~\mathrm{V})$ (solid circles), the $R$-mode splitting (open circles), and the intensity of the X-ray satellite Bragg peaks (open squares).}
\end{figure}

A close connection between the development of the electric order and the rapid increase of the resistance of the LAO/STO heterostructure is indeed observed as shown in Fig.~\ref{fig4}. Figures~\ref{fig4}(a) and \ref{fig4}(b) show that the voltage dependence of the resistance changes around 50~K. At higher temperature the resistance exhibits a weak, linear increase toward negative bias voltages. In contrast, at lower temperature the resistance increases much more rapidly and starts to exhibit a hysteretic behavior. Figure~\ref{fig4}(c) compares the voltage loops of the resistance at 10~K with the splitting of the $R$-mode which is a measure of the electric polarization. It confirms that the field-induced metal-to-insulator transition occurs in close vicinity to the pyroelectric transition. Notably, the resistance exhibits a similar hysteresis loop as the polarization of SrTiO$_3$. Finally, figure~\ref{fig4}(d) shows that the charge localization and the polar order develop below 50~K. This comparison suggests that the electric-field-induced transition of the confined electrons from a metallic or superconducting to a localized state is strongly influenced or likely even directly induced by the pyroelectric order.

In conclusion, we obtained evidence that the electric-field-induced localization of the confined electrons in LAO/STO heterostructures is induced, or at least strongly enhanced, by a pyroelectric phase transition which takes place in a thick ($\sim 1~\mu$m) SrTiO$_3$ layer adjacent to the interface. In particular, we showed that the charge localization and the polar order occur in the same voltage and temperature ranges and exhibit similar, unipolar hysteresis loops. Our findings suggest that this pyroelectric order and the subsequent reduction of the Coulomb screening also need to be considered for the interpretation of the quantum phase transition where superconductivity is suppressed by an electric field below $T_c\sim 0.2 - 0.4$~K \cite{Reyren2007}. It appears that these LAO/STO heterostructures are a unique platform where multiple long range orders, like superconductivity, ferromagnetism and now pyroelectricity, can be combined and readily modified with external parameters like electric or magnetic fields and possibly also with external pressure.


\begin{acknowledgements}
This work is supported by the Schweizer Nationalfonds (SNF) grants 200020-119784 and 200020-140225, by the NCCR MaNEP, by the German Science Foundation (TRR80), and the CEITEC by CZ.1.05/1.1.00/02.0068.  We appreciate the help of K. Conder and the support of Th. Lippert in preparing the isotope exchanged SrTi$^{18}$O$_3$ crystal. We acknowledge stimulating discussions with Ulrich Aschauer, Dionys Baeriswyl, Josef Humlicek, Andy Millis, Andrei Sirenko, Nicola Spaldin, and Jean-Marc Triscone.
\end{acknowledgements}

%


\newpage

\renewcommand{\thefigure}{S\arabic{figure}}
\setcounter{figure}{0}
\section*{Supplementary Online Material:\protect\\
Electric-field-induced pyroelectric order and localization of the confined electrons in L\lowercase{a}A\lowercase{l}O$_3$/S\lowercase{r}T\lowercase{i}O$_3$ heterostructures}

\section*{Infrared ellipsometry response of S\lowercase{r}T\lowercase{i}O$\mathbf{_3}$}

Figure S1 shows the infrared response of SrTiO$_3$ at representative temperatures of 10 and 300~K as measured with ellipsometry on a commercially available single crystalline substrate (Crystec). The data are shown in Fig. S1(a) in terms of the ellipsometric angles $\Psi$ and $\Delta$ at an incidence angle of 80$^{\circ}$ and in Fig. S1(b) of the derived real part of the optical conductivity. The black arrows mark the eigenfrequencies of the transverse-optical (TO) infrared-active phonon modes of the cubic phase at 300~K. These are the soft mode at $\sim95$~cm$^{-1}$\ (which decreases to $\sim15$~cm$^{-1}$\ at 10~K), the external mode at 175~cm$^{-1}$, and the bending mode at 544~cm$^{-1}$. The red arrow indicates the R-mode which develops below the anti-ferrodistortive phase transition at 105~K where the crystal structure becomes tetragonal. The temperature and electric field dependence of this $R$-mode is shown in detail in Figs. 1(d)-(g) and 2(c) and 2(d) of our paper. The grey arrow indicates the position of the highest longitudinal optical (LO) mode at $\sim860$~cm$^{-1}$. 

\begin{figure}[!ht]
  \centering
  \includegraphics[width=\columnwidth]{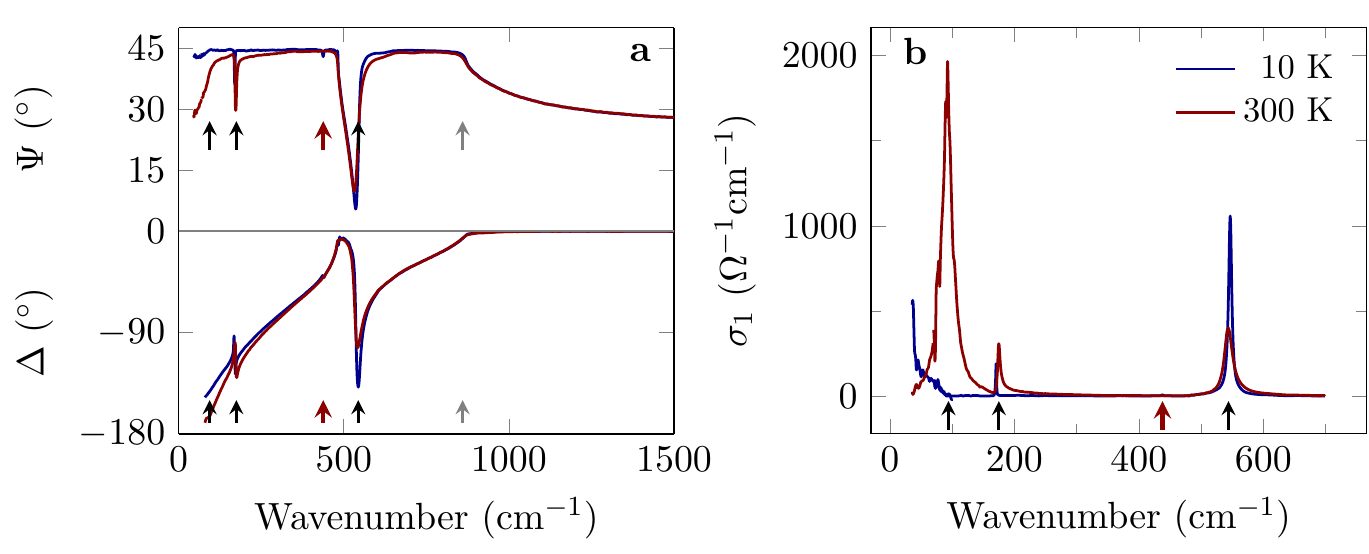}
  \caption{Ellipsometry spectra of a SrTi$^{16}$O$_3$ crystal at 10 and 300~K, respectively. \textbf{a}, Ellipsometric angles $\Psi$ and $\Delta$ and \textbf{b}, derived optical conductivity $\sigma_1$. Note the different energy ranges. The arrows indicate the phonon eigenfrequencies at 300~K (black arrows). The red arrow marks the R-mode which becomes weakly IR-active below 105~K. The LO edge of is marked by the grey arrow in \textbf{a}.}
  \label{fig:sto-spectra}
\end{figure}

\section*{Polarisation of S\lowercase{r}T\lowercase{i}O$\mathbf{_3}$ along the $\mathbf{[110]_c}$ or $\mathbf{[111]_c}$ axes}

In Fig. 1(c) of the paper we show a sketch of the ferroelectric polarisation due to a displacement for the Ti cations along the crystallographic $[001]_c$ direction (in cubic notation). Other possible Ti displacements are discussed in the following.

A shift of the Ti cation along the $[110]_c$ direction is schematically shown in Figure S2(a). Two of the Ti-O bonds are shortened and their eigenfrequency is blue-shifted here whereas four bonds are elongated and their eigenfrequencies are red-shifted as indicated by the colours of the bonds, respectively. Similar to the case of the $[001]_c$ displacement described in the paper, the $R$-mode thus also becomes anisotropic and splits into two branches corresponding to rotations and movements that involve either one blue and three red bonds or two blue and two red bonds, respectively.
\begin{figure}[!ht]
  \centering
  \includegraphics[width=\columnwidth]{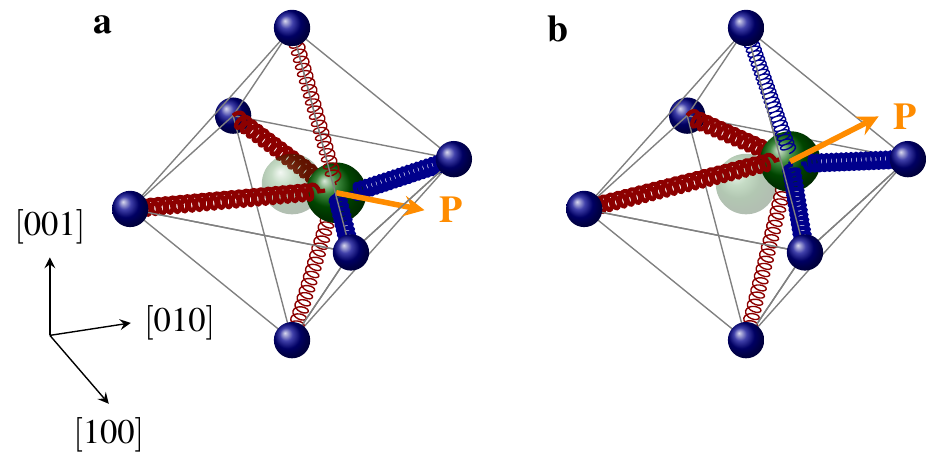}
  \caption{\textbf{a}, Sketch of the (exaggerated) polar displacement of the central Ti atom in a TiO$_6$ octahedron of STO along the $[110]_c$ axis and \textbf{b}, along the $[111]_c$ axis. The resulting polarisation is indicated by the orange vectors.}
  \label{fig:displacement}
\end{figure}

It has recently been argued in Ref. \cite{Shigenari2011} that such a $[110]_c$ displacement occurs in the ferroelectric state of SrTi$^{18}$O$_3$ for which the anomaly of the $R$-mode is shown in Figs. 2(a) and 2(b) of the paper. 

The third possible scenario is shown in Figure S2(b) where the Ti atom is displaced along the $[111]_c$ direction. This results in an equal number of three shorter and three longer bonds for which the $R$-mode remains isotropic. This case does not seem to be realised in SrTi$^{18}$O$_3$ nor in the LAO/STO heterostructures where the $R$-mode exhibits a clear splitting and therefore becomes anisotropic as is discussed in the manuscript. 

\section*{Modelling of the $\mathbf{R}$-mode at negative gate voltage}

For the fitting of the split $R$-mode at $V_g=-250$~V in Fig. 1(d) the following procedure has been applied. At first we have performed a fit of the spectrum at $+250$~V where only the $R$-mode due to the paraelectric SrTiO$_3$ is present. This fit was performed using three Lorentzian oscillators, one for the $R$-mode at 438~cm$^{-1}$\ and one each for the soft mode at very low frequency and the stretching mode at $\sim540$~cm$^{-1}$\ that are both well outside the spectral range that is shown in Fig. 1(d). The same modes and parameters were then subsequently used to account for the response of the paraelectric part of the SrTiO$_3$ substrate in the spectrum at $V_g=-250$~V. Here we introduced in addition a polar SrTiO$_3$ layer of thickness $d^{\textrm{polar}}$ at the LAO/STO interface in which the $R$-mode is softened and split into two peaks. For this additional layer we used the same parameters for the soft mode and the stretching mode, whereas the layer thickness, the eigenfrequencies, width and the spectral weight of the $R$-modes were fitted. The optical response due to the additional LAO layer on top of STO was found to be vanishingly small in the relevant spectral range between 410 and 460~cm$^{-1}$\ and therefore was neglected.

\section*{Field-induced anomaly of the soft mode of S\lowercase{r}T\lowercase{i}O$\mathbf{_3}$}

Using a home-built THz ellipsometer that is similar to the one discussed in Ref. \cite{Matsumoto2011}, we have investigated the field-dependence of the soft mode response of the LAO/STO heterostructure. In the paraelectric state the peak position of the soft mode decreases from $\sim95$~cm$^{-1}$\ at 300~K to $\sim15$~cm$^{-1}$\ at low temperature \cite{Cochran1960, Barker1962, Vogt1995}. In the pyroelectric state, the soft mode is expected to harden and to become anisotropic \cite{Yamanaka2000}. In the following we show that such a behaviour is indeed observed.

Figure S3 shows this effect in terms of the difference spectra of the ellipsometric angles, $\Psi(-250~\textrm{V}) - \Psi(+250~\textrm{V})$ and $\Delta(-250~\textrm{V}) - \Delta(+250~\textrm{V})$. The main features of the spectra are well reproduced by a model calculation (solid lines) for a 1~$\mu$m thick pyroelectric STO layer adjacent to the interface. The assumed hardening of the soft mode in the pyroelectric state at $-250$~V amounts to 1.8~cm$^{-1}$. In addition we introduced a regular, bipolar electric field-field-induced hardening of the soft mode of 2~cm$^{-1}$\ as it was previously observed in paraelectric SrTiO$_3$ crystals~\cite{Fleury1967, Fleury1968}
\begin{figure}[!ht]
  \centering
  \includegraphics[width=\columnwidth]{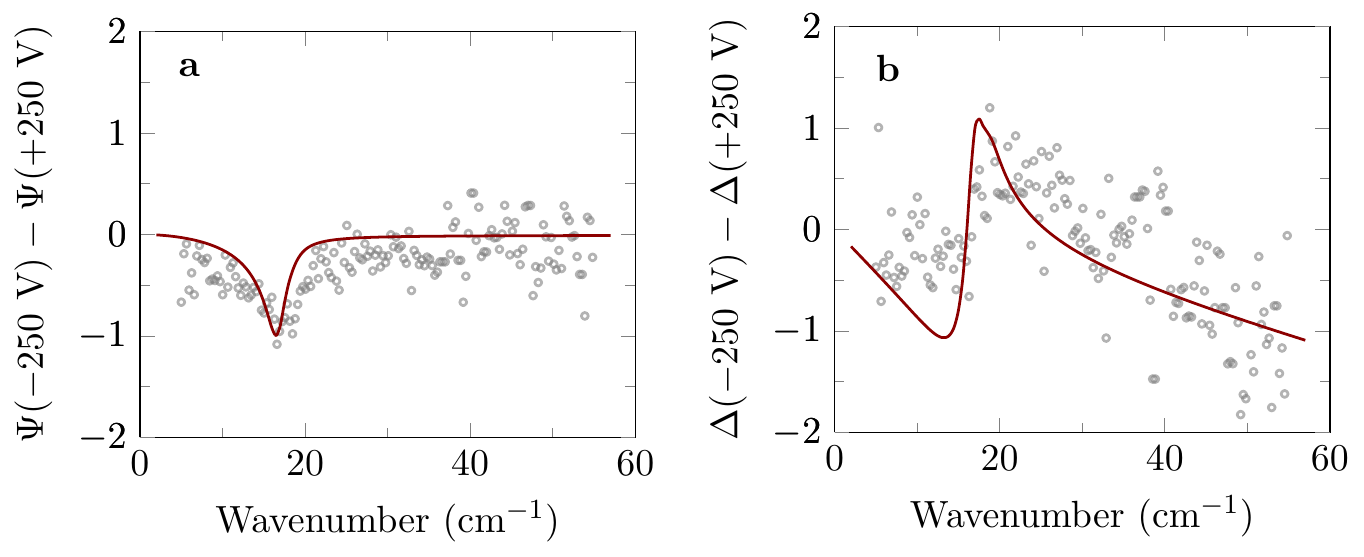}
  \caption{Difference spectra of the ellipsometric angles \textbf{a}, $\Psi(+250~\textrm{V}) - \Psi(-250~\textrm{V})$ and \textbf{b}, $\Delta(+250~\textrm{V}) - \Delta(-250~\textrm{V})$. The data of the LAO/STO heterostructure has been obtained at $T=10$~K. }
  \label{fig:r119-thz}
\end{figure}

Finally, a corresponding soft mode hardening in the ferroelectric phase has been observed in the SrTi$^{18}$O$_3$ crystal with $T^{\textrm{Curie}}=23$~K. Figure S4 shows the conductivity spectra derived from a Kramers-Kronig transformation of reflectivity data. These confirm that the soft mode hardens by $\sim5$~cm$^{-1}$\ between 30 and 5~K. Similar values have been previously obtained from Raman measurements on SrTi$^{18}$O$_3$ crystals \cite{Takesada2006}. 
\begin{figure}[!ht]
  \centering
  \includegraphics[width=\columnwidth]{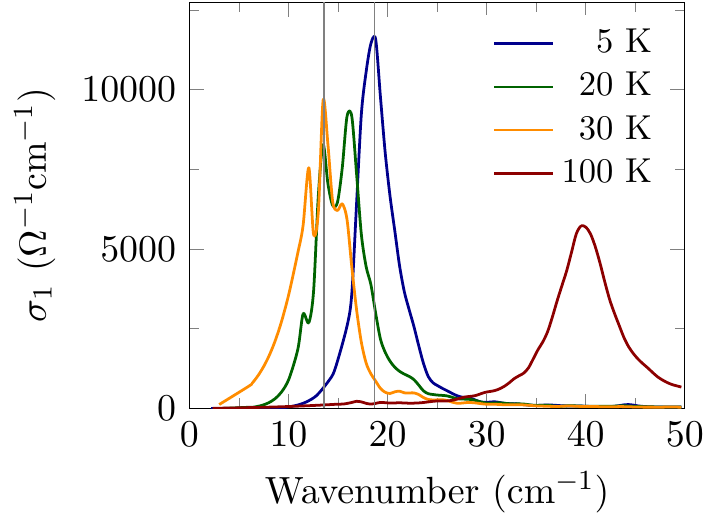}
  \caption{Optical conductivity spectra of the SrTi$^{18}$O$_3$ crystal showing the soft mode softening in the paraelectric phase down to 30~K and the subsequent hardening at temperatures below $T^{\textrm{Curie}}=23$~K.}
  \label{fig:sto18}
\end{figure}

\newpage

\end{document}